\documentclass[useAMS,usenatbib]{mn2e}

\usepackage{times,graphicx,amssymb,natbib,color}

\usepackage{amsmath}
\title[Kozai-Lidov cycles and Kepler-78b]{Can Kozai-Lidov cycles explain Kepler-78b?} 
\author[Ken Rice]{Ken Rice$^{1}$\thanks{E-mail: wkmr@roe.ac.uk}\\
$^{1}$Scottish Universities Physics Alliance (SUPA), Institute for Astronomy, University of Edinburgh, Blackford Hill, Edinburgh, EH9 3HJ}
\begin{document}

\date{Accepted 0000}

\pagerange{\pageref{firstpage}--\pageref{lastpage}} \pubyear{0000}

\maketitle

\label{firstpage}
\begin{abstract}
Kepler-78b is one of a growing sample of planets similar, in composition and size, to the Earth. It was first detected with NASA's \emph{Kepler} spacecraft and then characterised in more detail using 
radial velocity follow-up observations. Not only is its size very similar to that of the Earth ($1.2 R_\oplus$),
it also has a very similar density ($5.6$ g cm$^{-2}$). What makes this planet particularly
interesting is that it orbits its host star every $8.5$ hours, giving it an orbital distance of only $0.0089$ au. What we investigate
here is whether or not such a planet could have been perturbed into this orbit by an outer companion on an inclined
orbit.  In this scenario, the outer perturber causes the inner orbit to undergo Kozai-Lidov cycles which, if the periapse 
comes sufficiently close to the host star, can then lead to the planet being tidally circularised into a close orbit.  We find
that this process can indeed produce such very-close-in planets within the age of the host star ($\sim 600 - 900$ Myr), but it 
is more likely to find such ultra-short-period planets around slightly older stars ($> 1$ Gyr).  However, given the size of the {\em Kepler} sample and
the likely binarity, our results suggest that Kepler-78b may indeed have been perturbed into its current orbit by an outer stellar companion
The likelihood of this happening, however, is low enough that other processes - such as planet-planet scattering - could also be responsible.
\end{abstract}


\section{Introduction}
Analysis of data from NASA's \emph{Kepler} spacecraft \citep{borucki10,batalha13} indicates that planets
with radii similar to that of the Earth are common \citep{petigura13,dressing13}. 
Recently it was announced that one of the \emph{Kepler} targets (Kepler-78) showed a $0.02 \%$ decline in 
brightness that was associated with a planet with a radius of only $1.16 \pm 0.19 R_\oplus$ \citep{sanchis13}.
Follow-up observations, using HARPS-N \citep{cosentino12} and the High Resolution Echelle Spectrometer (HIRES) 
\citep{vogt94}, confirmed that this is indeed a planet with a mass of about $1.86 M_\oplus$ and a density
of about $5.6$ g cm$^{-3}$ \citep{pepe13,howard13}. 

Of course it is fascinating that we are now detecting planets with sizes and densities similar to that 
of the Earth, but what makes this planet particularly interesting is that it has an orbital period of only
8.5 hours, meaning that it is orbiting at a distance of only $0.0089$ au from its parent star. Quite how
such a planet can end up in such an orbit is very uncertain. It almost certainly could not have formed
where it now resides, as the temperature in the disc in that region would have been too high even for dust grains to
condense \citep{bell97}. It could potentially have migrated inwards through disc migration. However, such low-mass
planets would migrate in the gapless, Type I regime \citep{ward97} which is typically thought to
be so fast \citep{tanaka02,kley08} that it would seem unlikely that such objects could be left stranded
so close to their parent stars. Population synthesis models \citep{ida08,mordasini09} typically assume a 
reduced Type I migration rate.

Alternatively, such close-in planets could be scattered onto eccentric orbits \citep{ford06} that are then circularised
through tidal interactions with the parent star \citep{rasio96}. It
has indeed been suggested that if such a process were to operate, we should be seeing some very short period
hot super-Earths \citep{schlaufman10}, so this could be an explanation for the origin of Kepler-78b. However, even this
study suggested that typical orbital periods would be greater than the 8.5 hour orbital period of Kepler-78b.

Another mechanism for forming close-in planets, related to dynamical interactions in multi-planet systems, is
for the planet to undergo Kozai-Lidov cycles driven by a stellar companion on a highly inclined orbit 
\citep{kozai62,lidov62}.
If the eccentricity is sufficiently large, so that the periastron becomes very small, the planet's orbit may
be circularised through tidal interactions with its host star \citep{wu03,fabrycky07}. What we want to investigate
in this paper is whether or not this process could indeed explain the origin of Kepler-78b.  Given that binarity
amongst solar-like stars is quite high \citep{duquennoy91,abt06} it seems likely that this could play
a role in producing close-in, Earth-sized planets.

In this paper we present results from a series of Monte Carlo simulations in which we consider
how a planet with a mass and density the same as that of Kepler-78b, but initially orbiting between
0.5 and 2 au, is influenced by perturbations from a binary stellar companion. We also include
the influence of tides, which would allow the orbit to circularise if the eccentricity becomes
sufficiently large, and the influence of general relativistic and apsidal precession. The paper is organised
as follows : in Section 2 we present equations of motion, in Section 3 we describe the basic setup of
the problem, in Section 4 we discuss the results, and in section 5 we discuss the results and draw some 
conclusions.

\section{Equations of motion}
The goal is to evolve an inner binary (planet and star) under the influence of tidal 
interactions, perturbing accelerations from stellar and planetary distortions due to tides and 
rotation, perturbing accelerations from a third body, and general relativistic
apsidal precession.  To quadrupole order, these equations were first presented by 
\citep{eggleton01} and can also be found in \citet{wu03} and \citet{fabrycky07}.

Rather than using the equations in \citet{eggleton01}, we've implemented those from
\citet{barker09} and \citet{barker11} which are regular at $e = 0$.  We want to evolve an
inner system (planet + host star) where the bodies have masses $M_s$ and $M_p$, radii $R_s$ and $R_p$, 
and in which the orbit has an eccentricity $e$, 
semi-major axis $a$, and  orbital angular frequency $n = \sqrt{G (M_s + M_p)/a^3}$. 
The vector quantities that we want to evolve are, therefore, the spin of the
parent star ${\boldsymbol{\Omega_s}}$, the spin of the planet ${\boldsymbol{\Omega_p}}$, the eccentricity
of the inner orbit ${\boldsymbol{e}}$, and angular momentum vector of the inner orbit ${\boldsymbol{h} = 
\boldsymbol{r} \times \dot{\boldsymbol{r}}} = n a^2 \sqrt{1 - e^2} {\boldsymbol{\hat{h}}}$. 

We've built our model by considering, initially, only the equations that evolve
these quantities through tidal dissipation and a stellar wind.  From \citet{barker09} we have
\begin{align}
\begin{split}
\frac{d {\boldsymbol{h}}}{dt} &=  - \frac{1}{t_{fs}} \Bigg[ \frac{{\mathbf{\Omega_s} \cdot \boldsymbol{e}}}{2 n} f_5 (e^2) 
h {\boldsymbol{e}} \\
& - \frac{{\mathbf{\Omega_s}}}{2 n} f_3 (e^2) h + \Bigg( f_4 (e^2) - 
\frac{{\mathbf{\Omega_s} \cdot \boldsymbol{h}}}{2 n} \frac{1}{h} f_2 (e^2) \Bigg) {\boldsymbol{h}} \Bigg] \\
                     &  - \frac{1}{t_{fp}} \Bigg[ \frac{{\mathbf{\Omega_p} \cdot \boldsymbol{e}}}{2 n} f_5 (e^2)
h {\boldsymbol{e}} - \frac{{\mathbf{\Omega_p}}}{2 n} f_3 (e^2) h \\
& + \Bigg( f_4 (e^2) -
\frac{{\mathbf{\Omega_p} \cdot \boldsymbol{h}}}{2 n} \frac{1}{h} f_2 (e^2) \Bigg) {\boldsymbol{h}} \Bigg] \\
&= \Bigg( \frac{d {\boldsymbol{h}}}{dt} \Bigg)_s + \Bigg( \frac{d {\boldsymbol{h}}}{dt} \Bigg)_p
\label{eq:h1}
\end{split} \\
\begin{split}
h \frac{d {\boldsymbol{e}}}{dt} &=  - \frac{1}{t_{fs}} \Bigg[ \frac{{\mathbf{\Omega_s} \cdot \boldsymbol{e}}}{2 n} f_2 (e^2)
{\boldsymbol h} \\ 
& + 9 \Bigg( f_1 (e^2)h - 
\frac{11}{18}\frac{{\mathbf{\Omega_s} \cdot \boldsymbol{h}}}{n} f_2 (e^2) \Bigg) {\boldsymbol{e}} \Bigg] \\
                       & - \frac{1}{t_{fp}} \Bigg[ \frac{{\mathbf{\Omega_p} \cdot \boldsymbol{e}}}{2 n} f_2 (e^2)
{\boldsymbol{h}} \\ 
& + 9 \Bigg( f_1 (e^2) h -
\frac{11}{18}\frac{{\mathbf{\Omega_p} \cdot \boldsymbol{h}}}{n} f_2 (e^2) \Bigg) {\boldsymbol{e}} \Bigg] 
\label{eq:e}
\end{split} \\
\frac{d {\mathbf{\Omega_s}}}{dt} &= - \frac{\mu}{I_s} \Bigg( \frac{d {\boldsymbol{h}}}{dt} \Bigg)_s + \mathbf{\dot{\Omega}}_{s {\rm wind}}
\label{eq:Omegas} \\
\frac{d {\mathbf{\Omega_p}}}{dt} &= - \frac{\mu}{I_p} \Bigg( \frac{d {\boldsymbol{h}}}{dt} \Bigg)_p, 
\label{eq:Omegap}
\end{align}
where $I_s$ and $I_p$ are the moments of inertia of the star and planet, $\mathbf{\dot{\Omega}}_{s {\rm wind}}$  
represents the stellar wind, and $\mu = M_s M_p/(M_s + M_p)$ is the
reduced mass of the inner system.  We also need to define the tidal friction timescales for the star
and planet ($t_{fs}$ and $t_{fp}$), which depend on the star and planet's tidal quality factors ($Q'_{s}$
and $Q'_{p}$), and the functions of the eccentricity.

\begin{align}
\begin{split}
\frac{1}{t_{fs}} &= \Bigg( \frac{9 n}{2 Q'_{s}} \Bigg) \Bigg( \frac{M_{p}}{M_{s}} \Bigg) \Bigg( \frac{R_s}{a} \Bigg)^5 \\
\frac{1}{t_{fp}} &= \Bigg( \frac{9 n}{2 Q'_{p}} \Bigg) \Bigg( \frac{M_{s}}{M_{p}} \Bigg) \Bigg( \frac{R_p}{a} \Bigg)^5 
\label{eq:friction}
\end{split} \\
\end{align}
\begin{align}
f_1(e^2) &= \frac{1 + \frac{15}{4}e^2 + \frac{15}{8}e^4 + \frac{5}{64} e^6}{(1 - e^2)^\frac{13}{2}}
\label{eq:f1} \\
f_2(e^2) &= \frac{1 + \frac{3}{2}e^2 + \frac{1}{8}e^4}{(1 - e^2)^{5}}
\label{eq:f2} \\
f_3(e^2) &= \frac{1 + \frac{9}{2}e^2 + \frac{5}{8}e^4}{(1 - e^2)^{5}}
\label{eq:f3} \\
f_4(e^2) &= \frac{1 + \frac{15}{2}e^2 + \frac{45}{8}e^4 + \frac{5}{16} e^6}{(1 - e^2)^\frac{13}{2}}
\label{eq:f4}\\
f_5(e^2) &= \frac{3 + \frac{1}{2}e^2}{(1 - e^2)^5}
\label{eq:f5} \\
f_6(e^2) &= \frac{1 + \frac{31}{2}e^2 + \frac{255}{8}e^4 + \frac{5}{16} e^6 + \frac{25}{64} e^8}{(1 - e^2)^{8}}.
\label{eq:f6}
\end{align}
\noindent
We also include the contributions due to an additional outer body ($b$) of mass $M_o$, orbital angular frequency $n_o$, semi-major
axis $a_o$,  and
eccentricity $e_o$. Additionally, we add contributions from quadrupolar distortions of the
inner star and planet due to tidal and rotational bulges ($qs$ and $qp$), and we include general relativistic apsidal precession ($GR$).
The equations, shown below, are taken from \citet{barker11}

\begin{align}
\begin{split}
\Bigg( \frac{d {\boldsymbol{h}}}{dt} \Bigg)_b &= -3 C_b h \Bigg[ \frac{(1 - e^2)}{h^2} ({\boldsymbol{n \cdot h}})({\boldsymbol{n \times h}}) - \\
& \ \ \ \ \ \ \ \ 5 ({\boldsymbol{n \cdot e}})({\boldsymbol{n \times e}}) \Bigg]
\label{eq:hb} 
\end{split}\\
\begin{split}
\Bigg(\frac{d {\boldsymbol{h}}}{dt} \Bigg)_{qs} &= - \frac{\alpha_s}{h (1 - e^2)^2} ({\mathbf{\Omega_s} \cdot \boldsymbol{h}})
({\mathbf{\Omega_s} \times \boldsymbol{h}}) \\
\Bigg(\frac{d {\boldsymbol{h}}}{dt} \Bigg)_{qp} &= - \frac{\alpha_p}{h (1 - e^2)^2} ({\mathbf{\Omega_p} \cdot \boldsymbol{h}})
({\mathbf{\Omega_p} \times \boldsymbol{h}})
\label{eq:hq}
\end{split}\\
\begin{split}
h \Bigg( \frac{d {\boldsymbol{e}}}{dt} \Bigg)_b &= 3 C_b (1 - e^2) \big[2 ( {\boldsymbol{h \times e}} ) - ({\boldsymbol{n \cdot h}}) 
( {\boldsymbol{n \times e}} ) + \\ 
& \ \ \ \ \ \ \ 5 ({ \boldsymbol{n \cdot e}}) ({\boldsymbol{n \times h}}) \big]
\label{eq:eb}
\end{split}\\
\begin{split}
h \Bigg( \frac{d {\boldsymbol{e}}}{dt} \Bigg)_{qs} &=  \frac{\alpha_s}{(1 - e^2)^2} \Bigg[ \frac{1}{2} \Bigg( \frac{3}{h^2}
({\mathbf{\Omega_s} \cdot \boldsymbol{h}})^2 - \Omega_s^2 \Bigg) \\
& + \frac{15 G M_p}{a^3} f_2(e^2) (1 - e^2)^2 \Bigg] ({\boldsymbol{h \times e}}) \\
& + \frac{\alpha_s}{h^2 (1 - e^2)^2} ({\mathbf{\Omega_s} \cdot \boldsymbol{h}})({\mathbf{\Omega_s} \cdot \boldsymbol{h \times e}}){\boldsymbol{h}} \\
h \Bigg( \frac{d {\boldsymbol{e}}}{dt} \Bigg)_{qp} &=  \frac{\alpha_p}{(1 - e^2)^2} \Bigg[ \frac{1}{2} \Bigg( \frac{3}{h^2}
({\mathbf{\Omega_p} \cdot \boldsymbol{h}})^2 - \Omega_p^2 \Bigg) \\
& + \frac{15 G M_s}{a^3} f_2(e^2) (1 - e^2)^2 \Bigg] ({\boldsymbol{h \times e}}) \\
& + \frac{\alpha_p}{h^2 (1 - e^2)^2} ({\mathbf{\Omega_p} \cdot \boldsymbol{h}})({\mathbf{\Omega_p} \cdot \boldsymbol{h \times e}}){\boldsymbol{h}} 
\label{eq:eq}
\end{split}\\
h \Bigg( \frac{d {\boldsymbol{e}}}{dt} \Bigg)_{GR} &= \frac{3 G (M_s + M_p) n}{a c^2 (1 - e^2)} ({\boldsymbol{h \times e}} ),
\label{eq:eGR}
\end{align}
where $\boldsymbol{n}$ is a unit vector that is perpendicular to the plane of the outer body's orbit (not to be
confused with $n$ and $n_o$, which are the angular frequencies of the inner and outer orbits), and

\begin{equation}
\begin{split}
\alpha_s = \frac{R_s^5 k_s M_p}{2 \mu n a^5} \\
\alpha_p = \frac{R_p^5 k_p M_s}{2 \mu n a^5}
\end{split}
\label{eq:alpha}
\end{equation}
\noindent

\begin{equation}
C_b = \frac{M_o}{M_s + M_p + M_o} \frac{n_o^2}{n} \frac{1}{4 (1 - e^2)^{1/2} (1 - e_o^2)^{3/2}}.
\label{eq:Cb}
\end{equation} 
\noindent
In Equation (\ref{eq:alpha}), $k_s$ and $k_p$ are the inner star and planet's tidal love numbers.

\subsection{Octupole terms}
It now appears that expanding the equations only to quadrupole order may not be appropriate
for many systems \citep{naoz11,naoz12}, so we've also included the octupole terms. This allows
us to consider situations in which the outer body's mass is comparable to that of the inner planet,
and to consider situations in which the outer orbit is eccentric. 

We're unable to write the octupole terms in a way that is regular at $e = 0$, so have implemented
the form in \citep{mardling02}.  The octupole contributions are 

\begin{equation}
\begin{split}
\Bigg(\frac{d {\boldsymbol{h}}}{dt} \Bigg)_{oct} = & \frac{G (M_s + M_p)}{a} \Bigg( \frac{M_o}{M_s + M_p} \Bigg) 
\Bigg( \frac{M_s - M_p}{M_s + M_p} \Bigg)  \\  
& \times \Bigg( \frac{a}{R} \Bigg)^4 \frac{15 e}{16} \Bigg\{10 (1 - e^2)\hat{R_1}\hat{R_2}\hat{R_3}\hat{\boldsymbol{e}} \\
& + \Bigg[(4 + 3 e^2)\hat{R_3} - 5 ( 3 + 4 e^2) \hat{R_1}^2 \hat{R_3} \\
& - 5 (1 - e^2) \hat{R_2}^2 \hat{R_3} \Bigg] \hat{\boldsymbol{q}} \\
& - \Bigg[ (4 + 3 e^2) \hat{R_2} - 5 (1 + 6 e^2)\hat{R_1}^2 \hat{R_2} \\
& - 5 (1 - e^2) \hat{R_2}^3 \Bigg] \hat{\boldsymbol{h}}  \Bigg\}
\end{split}
\label{eq:hoct}
\end{equation} 

\begin{equation}
\begin{split}
\Bigg(\frac{d {\boldsymbol{e}}}{dt} \Bigg)_{oct} = & -n \Bigg( \frac{M_o}{M_s + M_p} \Bigg)
\Bigg( \frac{M_s - M_p}{M_s + M_p} \Bigg)  \\
& \times \Bigg( \frac{a}{R} \Bigg)^4 \sqrt{1 - e^2} \frac{15}{16} \Bigg\{ \Bigg[ -(4 + 3 e^2)\hat{R_2}^2 \\
& + (5 + 6 e^2) \hat{R_1}^2 \hat{R_2} + 5 ( 1 - e^2) \hat{R_3}^3 \Bigg] \hat{\boldsymbol{e}} \\
& + \Bigg[(4 + 3 e^2) \hat{R_1} - 5 (1 - 3 e^2) \hat{R_1} \hat{R_2}^2 \\
& - 5 (1 + 4 e^2) \hat{R_1}^3 \Bigg] \hat{\boldsymbol{q}} \\
& + 10 e^2 \hat{R_1} \hat{R_2} \hat{R_3} \hat{\boldsymbol{h}} \Bigg\},
\end{split}
\label{eq:eoct}
\end{equation}
where $\boldsymbol{R}$ is the co-ordinate of the outer body, and the co-ordinate frame is defined by 
the basis vectors ($\hat{\boldsymbol{e}}$,$\hat{\boldsymbol{q}}$,$\hat{\boldsymbol{h}}$), with $
\hat{\boldsymbol{q}} = \hat{\boldsymbol{h}} \times \hat{\boldsymbol{e}}$. The other unit vectors above
are $\hat{R_1} = \hat{\boldsymbol{R}}\cdot\hat{\boldsymbol{e}}$, $\hat{R_2} = \hat{\boldsymbol{R}}\cdot\hat{\boldsymbol{q}}$, 
$\hat{R_3} = \hat{\boldsymbol{R}}\cdot\hat{\boldsymbol{h}}$.

\subsection{Integrating the outer orbit}
The octupole terms described above need the co-ordinate of the outer body, which we determine by solving for
the eccentric anomaly, $E$.  This can be done by iterating the following equations until $dE$ is below
a threshold (we use $10^{-12}$)
\begin{equation}
\begin{split}
dE & = \frac{-(E - e_o\sin{E} - l)}{1 - e_o \cos{E}} \\
E & = E + dE,
\end{split}
\label{eq:ea}
\end{equation}
where $l$ is the mean anomaly $\big[ l = n_o (t - P) \big]$, with $P$ the orbital period and $t$ the time since the completion of the 
last full orbit of the outer body. In all of our simulations, we fix the outer body to lie in the $xy$ plane and 
so its co-ordinates are then
\begin{equation}
\begin{split}
R_x & = a_o (\cos{E} - e_o) \\
R_y & = a_o \sqrt{1 - e_o^2} \sin{E}\\
R_z & = 0,
\end{split}
\label{eq:outerorbit}
\end{equation}
where $a_o$ and $e_o$ are the semi-major axis and eccentricity of the outer orbit.  In this work, we neglect
perturbations on the outer orbit. 

\subsection{Stellar wind}
Without a stellar wind, or with a very weak stellar wind, it is possible that tidal interactions between the 
star and planet can result in the planet being trapped in a close orbit \citep{dobbs04}.  Most stars, however,
have winds that continue to remove angular momentum, and so once tidal interactions become significant, we 
would typically expect the planet to continue spiralling in towards the central star. We implement here
a very simple magnetic braking form for the stellar wind \citep{weber67,kawaler88,cameron94} so that
in the unsaturated regime, the stellar wind term is
\begin{equation}
{\dot{\Omega}}_{s\rm{wind}} = -\kappa_w {\Omega}_s^3,
\label{eq:wind1}
\end{equation}
where $\kappa_w$ is the braking efficiency coefficient. In the saturated regime (when $\Omega_s > \tilde{\Omega}$) this becomes
\begin{equation}
{\dot{\Omega}}_{s\rm{wind}} = -\kappa_w {\Omega}_s^2 \tilde{\Omega}.
\label{eq:wind2}
\end{equation}
The braking efficiency coefficient, $\kappa_w$, is set so that the stellar rotation period matches that expected
for the star being considered, and $\tilde{\Omega}$ is set to be $14 \Omega_\odot$.  The vector
associated with the stellar wind is always set so as to point in to opposite direction to that 
of the spin of the planet host star.  

\subsection{Putting it together}
Ultimately we want to evolve the angular momentum, $\boldsymbol{h}$, and eccentricity, $\boldsymbol{e}$, of the 
inner orbit, and the spins of the planet and its host star, $\mathbf{\Omega_p}$ and $\mathbf{\Omega_s}$. The evolution
of the stellar spin is determined by combining the stellar wind equations $\big[$Equations (\ref{eq:wind1}) and (\ref{eq:wind2}$\big]$
with Equation (\ref{eq:Omegas}).  The evolution of the spins of the star and planet $\big[$Equation (\ref{eq:Omegap})$\big]$, 
both depend on the tidal evolution of the 
orbital angular momentum $\big[$Equation (\ref{eq:h1})$\big]$.

To evolve the angular momentum of the inner orbit, we need to add the contributions from tides $\big[$Equation (\ref{eq:h1})$\big]$, 
perturbations from an outer body expanded to quadupole and octupole order $\big[$Equations (\ref{eq:hb}), and (\ref{eq:hoct})$\big]$, 
and perturbations from distortions of the inner star and planet $\big[$Equation (\ref{eq:hq})$\big]$. Similarly
to evolve the eccentricity of the inner orbit, we combine the contributions from tides $\big[$Equation (\ref{eq:e})$\big]$,
perturbations from an outer body expended to quadrupole and octupole order $\big[$Equations (\ref{eq:eb}), and (\ref{eq:eoct})$\big]$,
perturbations from distortions of the inner star and planet $\big[$Equation (\ref{eq:eq})$\big]$ and general
relativistic apsidal precession $\big[$Equation (\ref{eq:eGR})$\big]$.  

\subsection{Some basic tests}
Since this is a new code, we ran a few comparison tests to check that it was working properly.  The first was that
introduced by \citet{wu03}. It comprises a $1.1 M_\odot$ star with a $7.8 M_{\rm Jup}$ planetary companion, the star having a 
radius of $1 R_\odot$ and the planet having a radius the same as that of Jupiter. The initial stellar and planetary spin periods
are, respectively, 20 days and 10 hours.  The inner system's orbit has a semimajor 
axis of $a = 5.0$ au, and eccentricity $e = 0.1$, the tidal love numbers are $k_s = 0.028$
and $k_p = 0.51$, and the tidal dissipation quality factors are $Q'_s = 5.35 \times 10^{7}$ and $Q'_p = 5.88 \times 10^{5}$. The 
system also has a $1.1 M_\odot$ companion with $a_o = 1000$ au,  $e_o = 0.5$ and with an orbital plane inclined at $85.6^o$ to that
of the plane of the inner orbit.

\begin{figure}
\begin{center}
\includegraphics[scale = 0.45]{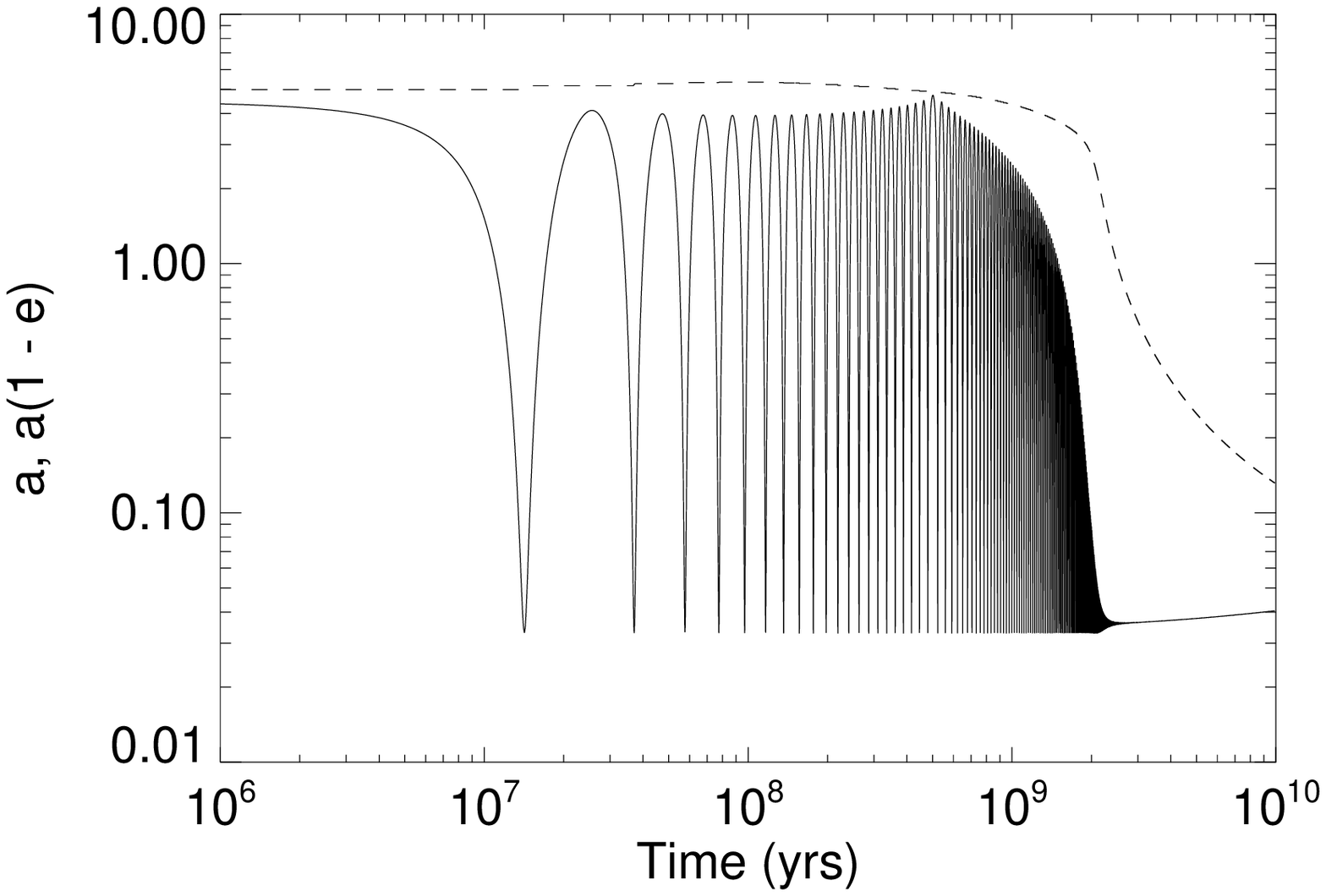}
\caption{A figure showing the evolution of the semimajor axis ($a$ - solid line) and the periaps $\left(a(1-e) - {\rm dashed \  line} \right)$ using initial
conditions the same as those in \citet{wu03}. This was a code test that was also carried out by \citet{fabrycky07}, and our
results appear to match theirs.  It doesn't quite match \citet{wu03} but we can match their results if we
ignore the term representing the apsidal precession due to the spin and tidal bulges of the planet.}
\label{fig:wumu03}
\end{center}
\end{figure}

Figure \ref{fig:wumu03} shows the time evolution of the semi-major axis (dashed line) and periaps (solid line) of the system described above, 
and appears the same as that in \citet{fabrycky07}, who also performed this test.  It's not
quite the same as in \citet{wu03}, but we can match their result if we remove the apsidal precession due to the spin and tidal bulges of
the planet (which can require very short timesteps and, hence, long integration times).  

The second test was primarily to check that the octupole terms had been properly implemented.  In this test, taken from \citet{naoz13}, 
we ignore the tidal evolution terms,
the terms associated with the distortion of the inner star and planet due to their tidal bulges, and the effect of general relativistic apsidal
precession.  The system consists of an inner star of mass $1 M_\odot$, a companion planet with mass $1 M_{\rm Jup}$, and an outer planet with mass
$1 M_\odot$.  The inner orbit has a semimajor axis of $a = 6$ au and eccentricity of $e = 0.001$.  The outer orbit has a semimajor axis of
$a_o = 100$ au, an eccentricity of $e_o = 0.6$ and is inclined at $65^o$ to the plane of the inner orbit.  The argument of pericentre
of the inner orbit is also set, initially, to $45^o$ with the outer one set to zero.

\begin{figure}
\begin{center}
\includegraphics[scale = 0.45]{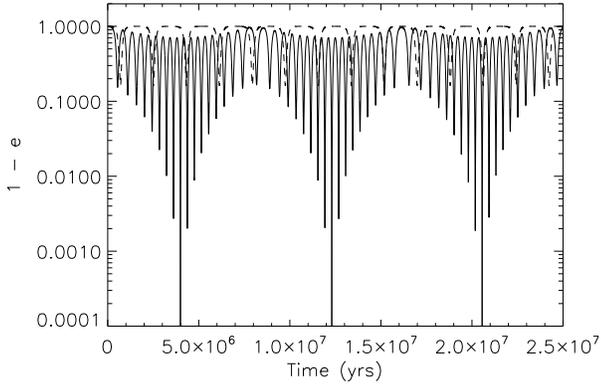}
\caption{The evolution of $1 - e$ for a system with the same parameters as those used by \citet{naoz13} and described
here in the text.  The solid line shows the evolution when the octupole terms are included and it matches that of \citet{naoz13}. 
The dashed line shows how the system would evolve in the absence of the octupole terms.}
\label{fig:naoz13}
\end{center}
\end{figure}

Figure \ref{fig:naoz13} shows the time evolution of $1 - e$ for the system described above.  The result appears identical
to that in \citet{naoz13}.  The dotted line also shows how the system would evolve in the absence of the octupole terms.

\section{Basic setup}
The system we want to consider specifically is Kepler-78 \citep{sanchis13}.  The companion planet, with a
mass of $M_p = 1.86 M_\oplus$ and radius of $R_p = 1.173 R_\oplus$, is 
Earth-sized and has an Earth-like density ($\rho = 5.57$ g cm$^{-3}$) \citep{pepe13,howard13}. The host star
has a mass of $M_s = 0.81 M_\odot$ and radius $R_s = 0.737 M_\odot$, and the planet has an orbit that is circular ($e = 0$) and
orbits at a distance $a = 0.0089$ au (giving an orbital period of 8.5 hours).  

The system is thought to have an age between $600$ and $900$ Myr \citep{sanchis13}, so here we run our initial simulations for $800$ Myr.  We set
the stellar wind braking parameter to $\kappa_w = 10^{46}$, which gives a stellar rotation speed of between $11$ and $12$ days at $t = 800$ Myr, similar
to that observed \citep{pepe13}. A star with a mass similar to that of Kepler-78, however, does not spin down much in the first Gyr and so the 
stellar wind is probably not particular important here.  The tidal love numbers are set to $k_s = 0.028$ and $k_p = 0.51$, and we consider stellar 
tidal quality factors of $Q'_s = 5 \times 10^5$ and $Q'_s = 5 \times 10^6$.  
We'll specify the tidal quality factor for the planet $Q'_p$, the planet's initial orbital 
properties and the properties of the outer body, when we discuss the results of the simulations. 

\section{Results}
Since we want to consider if Kozai-Lidov cycles could explain the properties of Kepler-78b, our models are set up in the 
following way.  We assume we have a planet with a mass and radius the same as that of Kepler-78b with an initial semi-major axis
between $a = 0.5$ au and $a = 2$ au, with the semi-major axis chosen randomly in $\log a$.  The initial eccentricity is set to be $e = 0.05$, chosen
because we're assuming, here, that the planet has formed in a circumstellar disc in an almost circular orbit.  We should acknowledge, however,
that the initial eccentricity can have a significant impact on the evolution of co-planar systems \citep{li14} and, therefore, stress that
our results only apply to systems in which the inner system has a low initial eccentricity.

We also assume that there is an outer companion
with a mass randomly chosen to be uniform between $M_o = 0.1 M_\odot$ and $M_o = 1 M_\odot$, a semi-major axis chosen randomly in $\log a$,
between $a_o = 40$ au and $a_o = 20000$ au, and a randomly chosen eccentricity between $e_o = 0$ and $e_o = 1$.  We then fix the 
outer companion's orbit to be in the $xy$ plane and randomly orientate the inner orbit so that the mutual
inclination, $i$, is isotropic \citep{wu07}. We also randomly orientate the longitude of the planet's ascending node. By choosing such 
a high-mass companion, we're essentially in the test particle regime \citep{lithwick11}.  Such companions will also
produce a large maximum eccentricity (for the inner orbit) than lower mass companions \citep{teyssandier13}. As such, we might expect 
a reasonably large number of tidal disruption events \citep{naoz12, li14, petrovich15}.  As such, our results only apply to a situation where
the companion is of stellar mass. 

We also impose stability critera \citep{lithwick11,naoz13,mardling01} and inisist that
\begin{equation}
\frac{a}{a_o} \frac{e_o}{1 - e_o^2} < 0.1,
\label{eq:stab1}
\end{equation}
and that
\begin{equation}
\frac{a_o}{a} > 2.8 \left(1 + \frac{M_o}{M_s + M_p} \right)^{2/5} \frac{(1 + e_o)^{2/5}}{(1 - e_o)^{6/5}} \left(1 - \frac{0.3 i}{180^o} \right).
\label{eq:stab2}
\end{equation}
Equation (\ref{eq:stab1}) ensures that we are in the regime where the quadrupole and octupole terms dominate, while Equation (\ref{eq:stab2}),
in which $i$ is the mutual inclination of the two orbits, ensures that the triple system is long-term stable \citep{mardling01}.
Equation (\ref{eq:stab2}) is almost always satisifed for the initial conditions used here. 

\subsection{Initial results}
\begin{figure}
\begin{center}
\includegraphics[scale = 0.45]{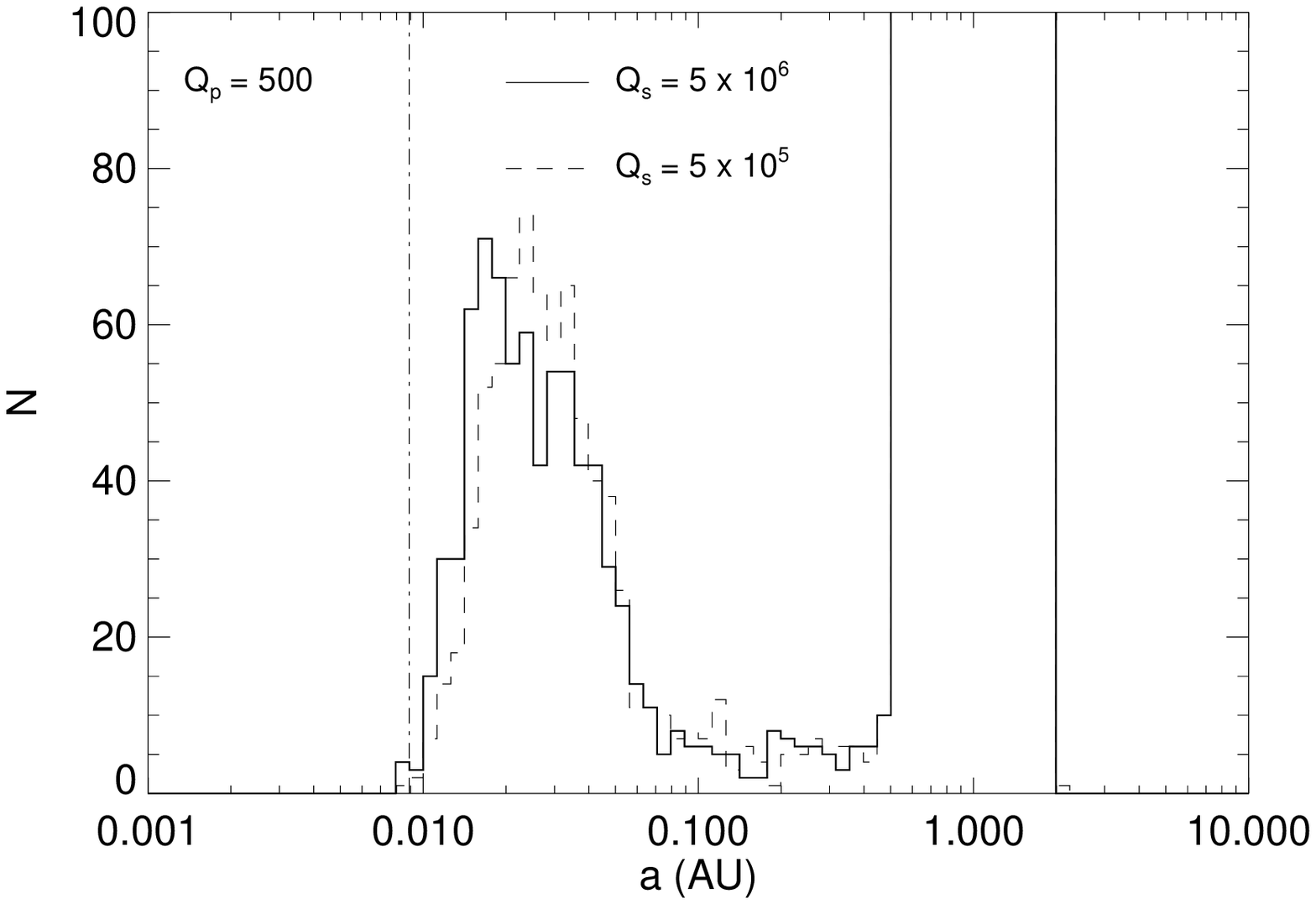}
\includegraphics[scale = 0.45]{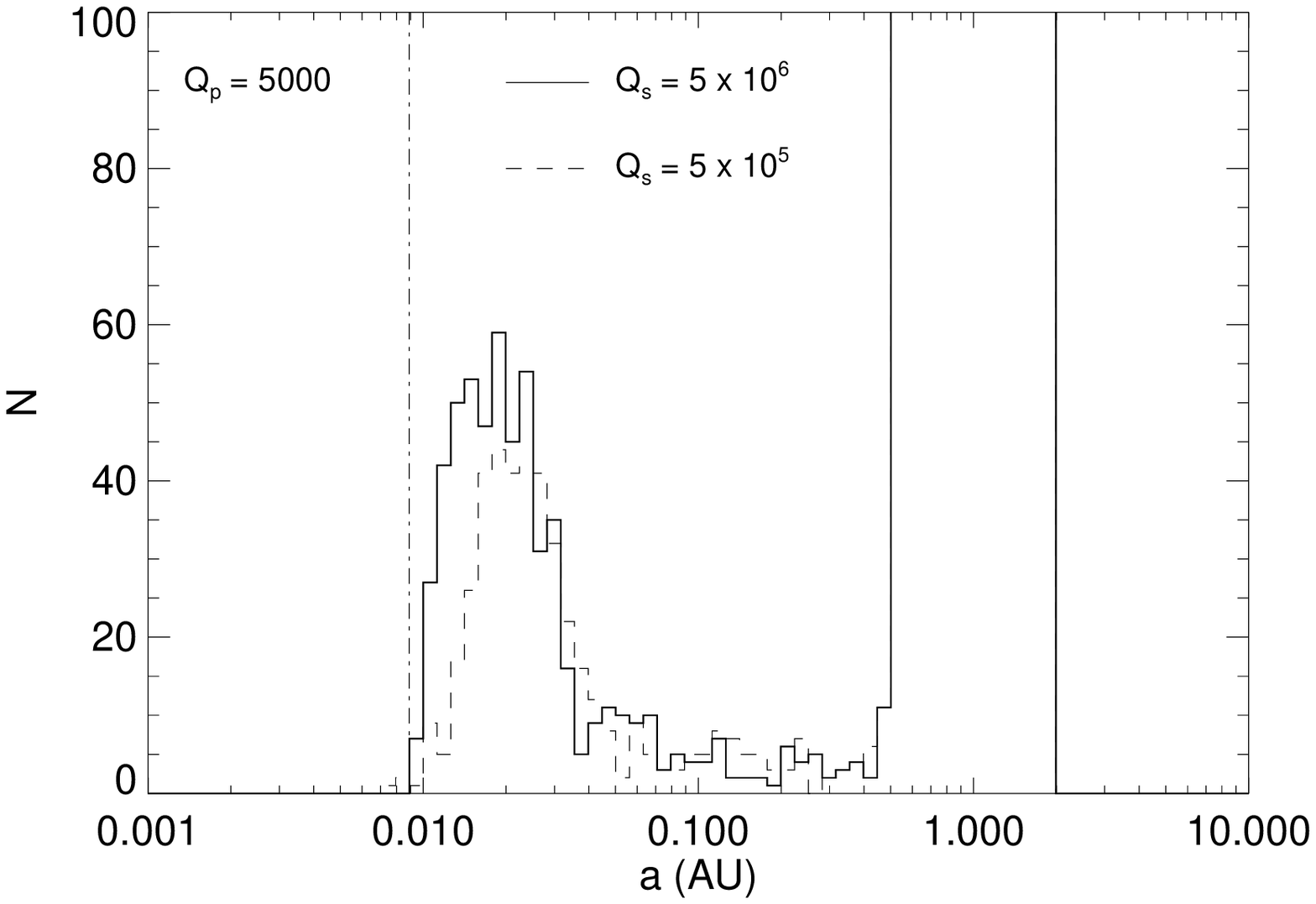}
\caption{Histograms showing the final semi-major axes of the simulations with initial setup as described in the 
text and in which each simulations is stopped at $t = 800$ Myr. 
The top panel is for $Q'_p = 500$ and the bottom is $Q'_p = 5000$.  The stellar tidal qualifty factors that
we consider are $Q'_s = 5 \times 10^6$ (solid line) and $Q'_s = 5 \times 10^5$ (dashed line).  We consider 10000 system in each case, with the 
planet starting with $a$ between $0.5$ au and $2$ au.  After $800$ Myr, there are between $10$ and $4$ planets surviving inside $a = 0.01$ au, depending on
the values of $Q'_p$ and $Q'_s$. The vertical dash-dot line indicates the current semimajor axis of Kepler-78b. In each case, the number of planets
still located between $0.5$ and $2$ au is very large, and their distribution extends well above the limits shown on the y-axis.}
\label{fig:a_final_log_tmax8d8}
\end{center}
\end{figure}

The tidal quality factor for a terrestrial planet is thought to lie between $Q'_p = 10$ and $Q'_p = 500$ \citep{goldreich66}. Since
we're considering a young system in which the planet likely retains a lot of its initial internal heat, we assume 
a value at the top of this range ($Q'_p = 500$), and also a more extreme case where $Q'_p = 5000$ \citep{henning09}.  For the star, we assume tidal
quality factors of $Q'_s = 5 \times 10^5$ and $Q'_s = 5 \times 10^6$, within the range expected for exoplanet host stars
\citep{baraffe10,brown11}.  For each simulation we select the initial conditions as described above and evolve the 
system until $t = 800$ Myr, using a fourth-order Runge-Kutta integrator.  We repeat this 10000 times for each set of parameters, 
and the basic result is shown in Figure \ref{fig:a_final_log_tmax8d8}.  The top panel is for $Q'_p = 500$ and the bottom for $Q'_p = 5000$. The solid line in each figure is for $Q'_s = 5 \times 10^6$, the dashed line is for $Q'_s = 5 \times 10^5$, and the
vertical dash-dot line indicates the current semimajor axis of Kepler-78b.  In each case, the number of planets still located between $0.5$ and $2$
is very large and their distribution extends well above the limits shown on the y-axis. 

From Figure \ref{fig:a_final_log_tmax8d8} it seems clear that it is possible for a planet to be perturbed into an orbit
inside $a = 0.01$ au within 800 Myr.  However, the numbers are typically small.  For $Q'_p = 500$ it is $10$ ($Q'_s = 5 \times 10^6$)
and $4$ ($Q'_s = 5 \times 10^5$), while for $Q'_p = 5000$ it is $7$ and $4$ respectively.
Even though the number of planets surviving inside $a = 0.01$ au is small, in most cases, a much larger number
reach their Roche limit [$a = R_p/0.462 (M_*/M_p)^{1/3} = 0.0056$ au] \citep{faber05} and are assumed to be tidally disrupted and
destroyed.  With the exception of the $Q'_p = 500$, $Q'_s = 5 \times 10^6$ simulation (in which the numbers were small), 
in excess of 100 - out of a sample of 10000 - reached the Roche limit.

\begin{figure} 
\begin{center}
\includegraphics[scale = 0.45]{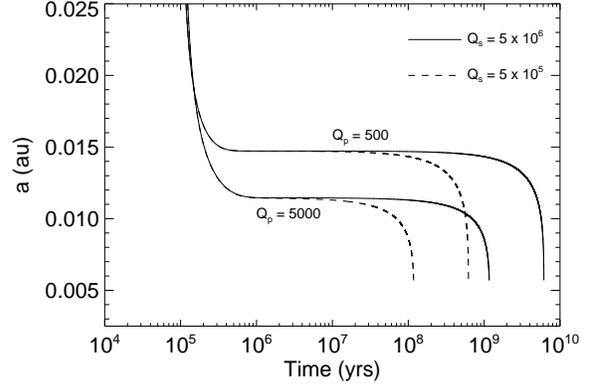}
\caption{A set of single planet simulations, each with a different $Q'_p$ or $Q'_s$ value.  In each case we've chosen inital
conditions that would indeed perturb the planet into a close orbit, and have run each simulation until the planet reaches its
Roche limit ($a = 0.0056$ au in this case). The time a planet spends inside $0.01$ au before reaching its Roche limit depends 
primarily on the star's tidal quality factor and varies from $48$ Myr ($Q'_s = 5 \times 10^5$) to $480$ Myr ($Q'_s = 5 \times 10^6$).}  
\label{fig:singleplanet}
\end{center}
\end{figure}

Given that a large numbers of planets do become tidally destroyed, it is useful to know for how long a planet might exist inside $a = 0.01$ au. 
Figure \ref{fig:singleplanet} shows four single planet simulations, one for each combination of $Q'_p$ and $Q'_s$, each of which is
run until the planet reaches its Roche limit. The
amount of time such a planet spends inside $a = 0.01$ au depends, primarily, on the star's tidal quality factor. For $Q'_s = 5 \times 10^6$,
the planet reaches the Roche limit in 480 Myr, while for $Q'_s = 5 \times 10^5$ it takes 48 Myr. Therefore, it would seem that
for reasonable estimates of the star's tidal quality factor, a planet such as Kepler-78b will only be detectable inside $a = 0.01$ au
for a few hundred Myrs at most.
 
Our initial results would therefore seem to suggest that it is possible for an outer companion to perturb a planet
like Kepler-78b into a very close orbit ($a < 0.01$ au) within the age of the system ($\sim 800$ Myr). However, the 
numbers are small, with at most $10$ out of $10000$ surviving inside $a = 0.01$ au at $t = 800$ Myr.

\subsection{Age of the system}
The previous simulations only considered the likelihood of a system with an age similar to that of Kepler-78,
having a planetary companion with orbital properties similar to the of Kepler-78b.  The results suggest
it is possible, but probably rare.  Additionally, the age distribution of a sample of 950 {\em Kepler} object 
of interest host stars \citep{walkowicz13} suggests that about 10\% have ages less than 1 Gyr.  This is also
probably biased towards younger stars because of the way in which the sample was selected. Given that about 
40\% of these would have stellar companions \citep{raghavan10} would further reduce the likelihood of actually observing a Kepler-78b
type system.

\begin{figure}
\begin{center}
\includegraphics[scale = 0.45]{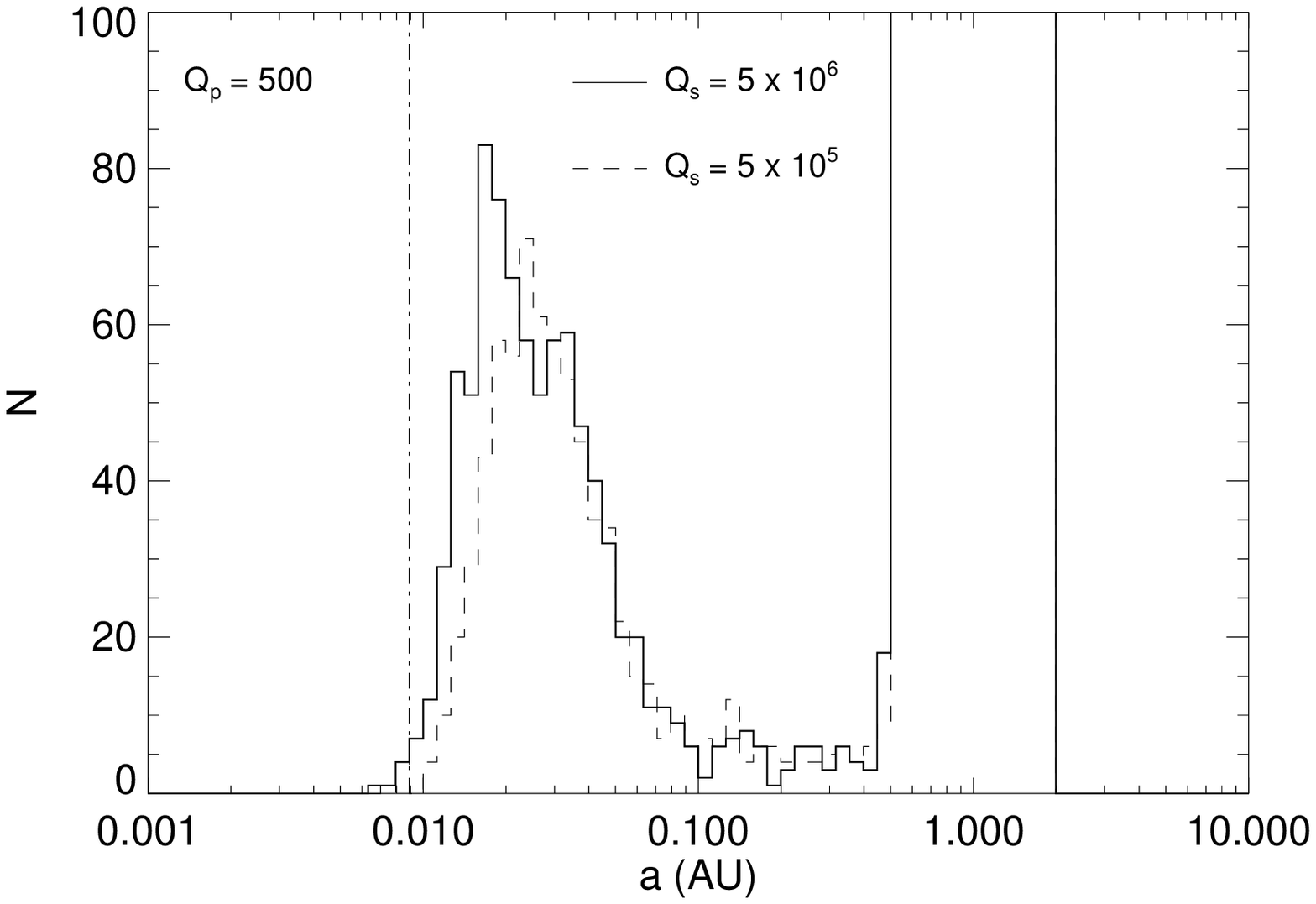}
\includegraphics[scale = 0.45]{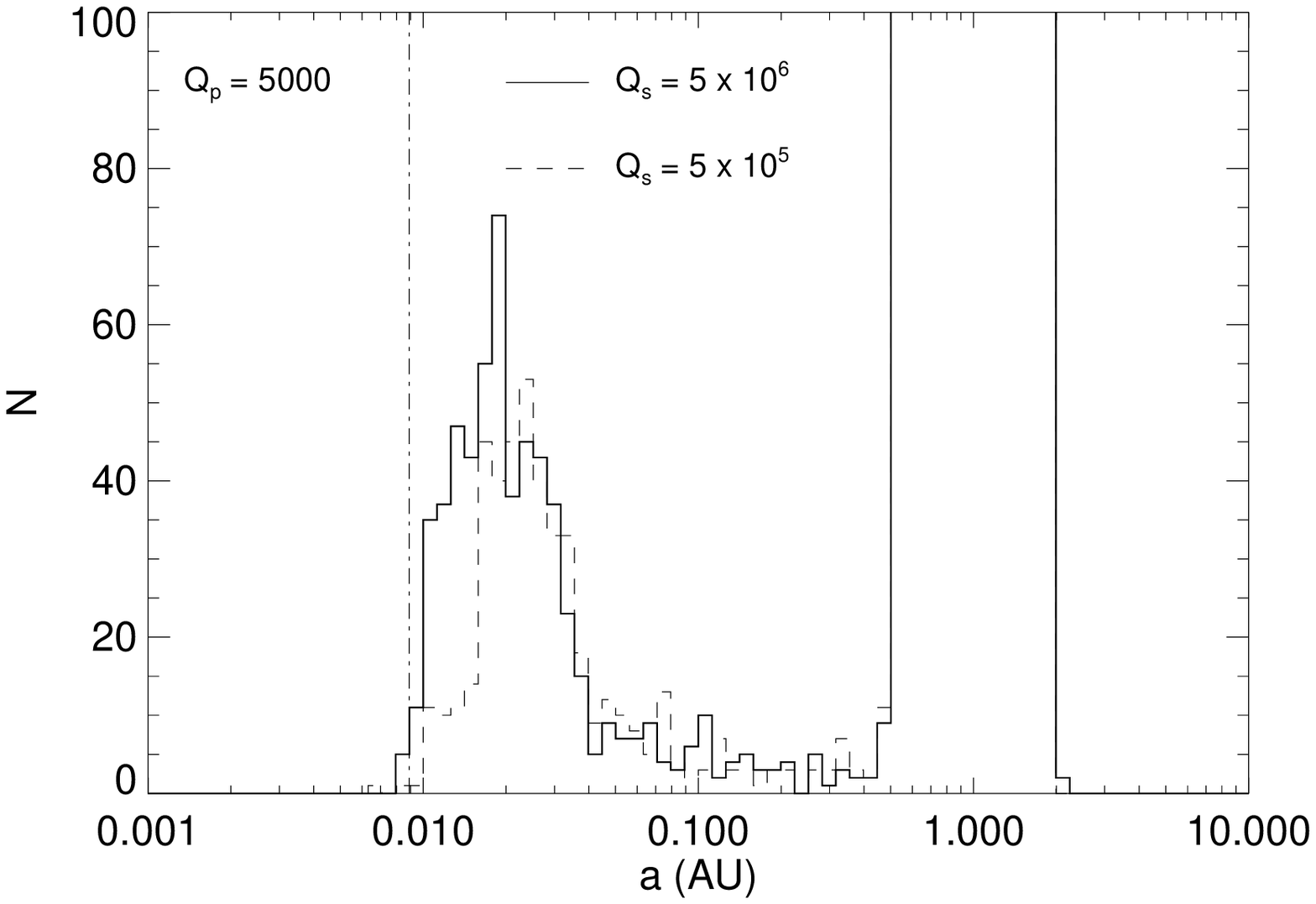}
\caption{Histograms showing the final semi-major axes for the simulations with initial setup as described in the 
text and in which the age of each system is randomly chosen to be between $500$ Myr and $2$ Gyr.
The top panel is for $Q'_p = 500$ and the bottom is $Q'_p = 5000$.  The stellar tidal qualifty factors that
we consider are $Q'_s = 5 \times 10^6$ (solid line) and $Q'_s = 5 \times 10^5$ (dashed line).  We consider 10000 system in each case, and the
planets start with $a$ between $0.5$ and $2$ au. In each case, the number of planets
still located between $0.5$ and $2$ au is very large, and their distribution extends well above the limits shown on the y-axis. Depending on the 
values of $Q'_p$ and $Q'_s$, there are between $17$ and $6$ planets surviving inside $a = 0.01$ au.}
\label{fig:a_final_log_tmax2d9}
\end{center}
\end{figure}

To investigate how our results might depend on the age of the system, we reran our simulations with
the same set of parameters as described above, but allowed the age of the star to vary, uniformly, from $500$ Myr, to $2$
Gyr. The resulting histograms are shown in Figure \ref{fig:a_final_log_tmax2d9} and are very similar to those in
Figure \ref{fig:a_final_log_tmax8d8}.  There is a slight increase in the number surviving inside $a = 0.01$ au for 
$Q'_s = 5 \times 10^6$. In these runs there were 17 and 16 for $Q'_p = 500$ and $Q'_p = 5000$ respectively, compared to $10$ and $7$ 
when the age of the system was fixed at $t = 800$ Myr. For $Q'_s = 5 \times 10^5$, the numbers are similar to the runs with 
the age fixed at $t = 800$ Myr.
  
\begin{figure}
\begin{center}
\includegraphics[scale = 0.45]{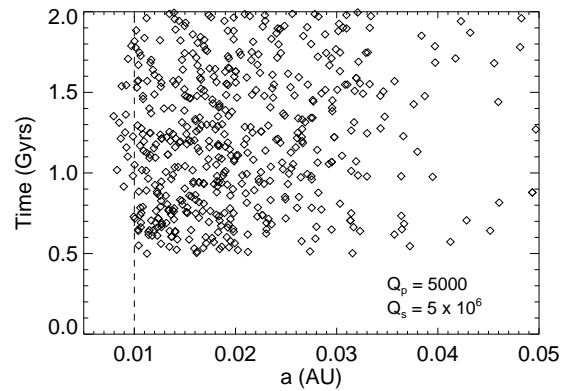}
\caption{Figure showing the final semimajor axes for those planets that end up with $a < 0.05$ au, plotted against the age of
the system.   This simulations each considered 10000 systems in which the outer perturber was assumed to 
have a semimajor axis distribution that extended to $a_o = 20000$ au.
It's clear that it is possible for a system with an age similar to that of Kepler-78 to have a planet inside $0.01$ au, but it is more
likely for systems older than $1$ Gyr, than for those with ages below $1$ Gyr.}
\label{fig:apl_time}
\end{center}
\end{figure}

To further see the influence of the age of the system, we plot in Figure \ref{fig:apl_time} the final semimajor axis of the planet
against age of the system, for all those systems in which planets end up inside $0.05$ au.  We only show, however, the results for
$Q'_p = 5000$, $Q'_s = 5 \times 10^6$ as that produced the largest number of surviving planets inside $a = 0.01$ au.  The first thing
to note is that it appears more likely to detect such a planet for systems older than 1 Gyr (Kepler-78 has an age of between
$600$ and $900$ Myr). However, Figure \ref{fig:apl_time} does show 4 systems with an age $< 1$ Gyr, and with a planet inside $a = 0.01$ au.  

\citet{walkowicz13} also suggest that maybe 20\% of the {\em Kepler} targets have ages less than $2$ Gyr.  \emph{Kepler} observed 
about 150000 stars \citep{borucki10}, which suggests maybe as many as many as 30000 could have ages less than $2$ Gyr.
Candidates as small as Kepler-78b, however, are typically found around quieter - and therefore older - stars.
\emph{Kepler} is therefore incomplete for stars with high Combined Differential Photometric Precision \citep{batalha13,
christiansen13} and so the number of such planets is likely an underestimate.
In the scenario shown in Figure \ref{fig:apl_time}, $17$ - out of $10000$ - survive inside $a = 0.01$ au. If we assume that 
$40$\% of those stars have stellar companions \citep{raghavan10}, and that all of those stars could 
host terrestrial planets \citep{cassan12,greaves11}, then we might expect as many as $20$ of the $30000$ 
{\em Kepler} targets, with ages below $2$ Gyr, to host such a ultra-close-in planet. Of course, our other simulations suggest that
the number surviving could be as low as $2$ (depending on the tidal properties of the star and planet), but given that the chance of such a 
system transiting is actually quite high ($46 \%$), observing such a system is still quite likely. 

Similarly, if we consider only those systems with ages below $1$ Gyr, Figure \ref{fig:apl_time} suggests that maybe as many as $4$ out of 
$10000$ could survive inside $a = 0.01$ au. Repeating the calculation above suggests that maybe $4 - 5$ stars with ages similar to that
of Kepler-78 could host such a close-in planet.  Again, given the high transit probability for
such a close-in system, detecting a planet such as Kepler-78b becomes possible. Our results therefore suggest that it is possible
for this process to have produced a planet like Kepler-78b. Of course, if Kepler-78 is closer in age to 
$600$ Myr, than to $900$ Myr, Figure \ref{fig:apl_time} suggests that it would become less likely.

\subsection{Perturber properties}
\begin{figure}
\begin{center}
\includegraphics[scale = 0.45]{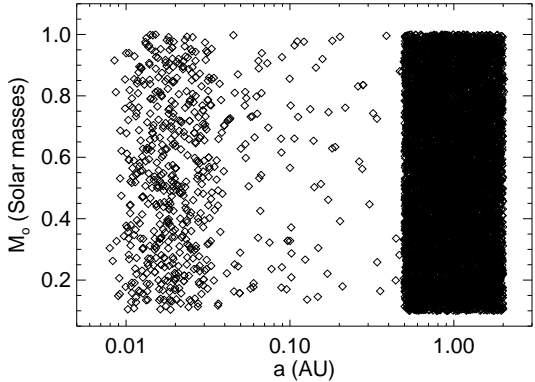}
\caption{Figure showing the planet's final semi-major axis plotted against the
mass of the outer companion. There appears to be little dependence on companion mass. \emph{Kepler} would probably have detected a 
companion with a mass in excess of $\sim 0.5$ M$_\odot$, but this figure does show that a low-mass companion ($M < 0.5$ M$_\odot$)
could indeed have produced a system like Kepler-78b system.}
\label{fig:mass_a}
\end{center}
\end{figure}

The results above suggest that it is possible for an outer perturber to drive a Kepler-78b-like planet into a close-in orbit within the
age of Kepler-78. To see how the properties of the outer body influences the 
inner planet, we show - in Figure \ref{fig:mass_a} - how
the final semimajor axis of the planet depends on the mass of the outer body.  Again, we only shows results from the 
simulation with $Q'_p = 5000$ and $Q'_s = 5 \times 10^6$. Figure \ref{fig:mass_a} suggests 
that there isn't a particularly
strong mass dependence, consistent with our simulations essentially being in the test particle regime \citep{lithwick11}. 
However, Kepler-78, which has an apparent magnitude of $m_v = 12$,  is not known to host a stellar companion.
Since \emph{Kepler} is sensitive down to an apparent magnitude of $m_v = 14$, that would suggest that if there is an undetected 
companion it would need to have a mass less that about $0.5$ M$_\odot$. 

\begin{figure}
\begin{center}
\includegraphics[scale = 0.45]{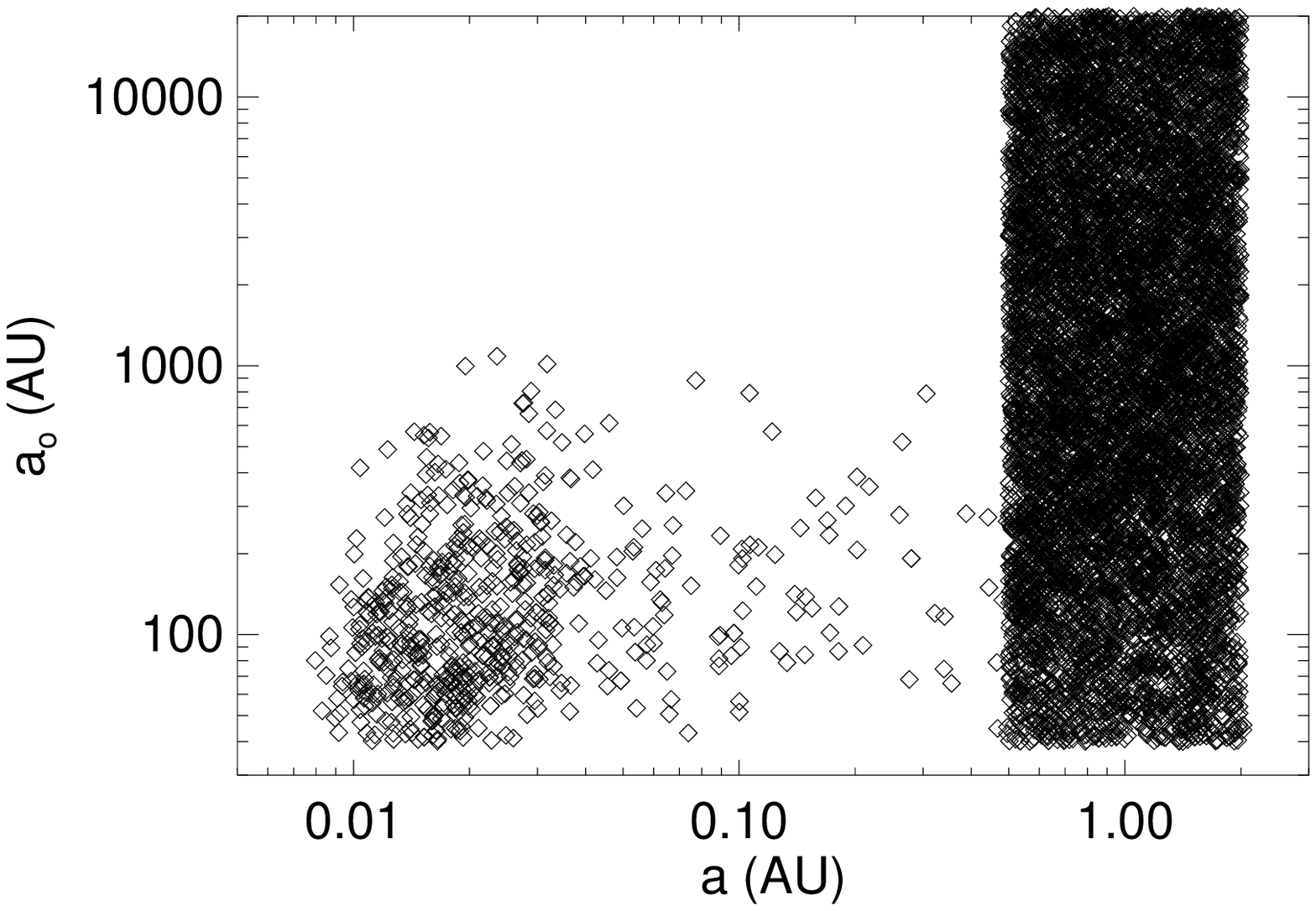}
\includegraphics[scale = 0.45]{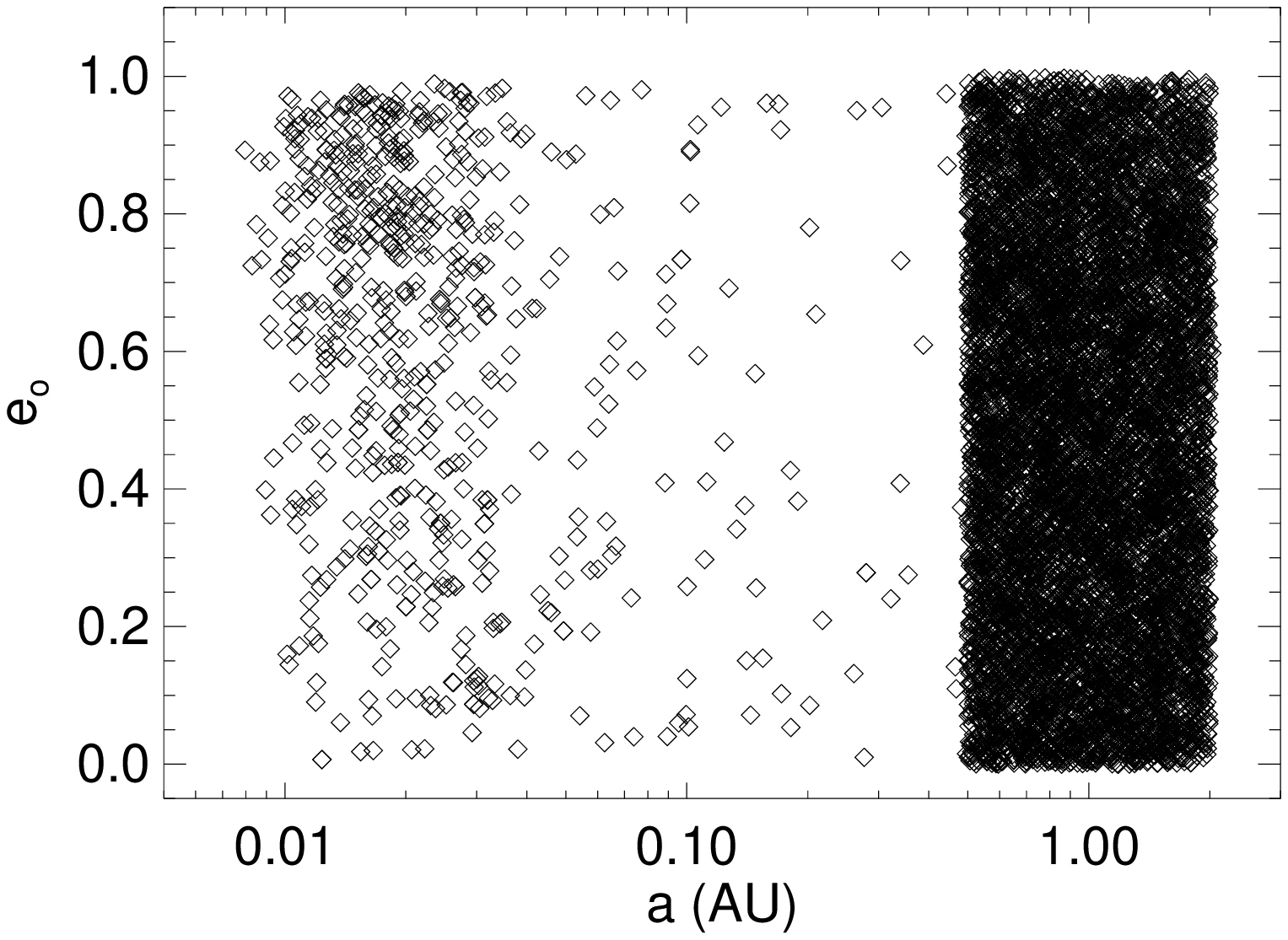}
\caption{Figure showing how the orbital properties of the outer companion influences the final semimajor axis of
the inner planet.  The top panel shows that outer companions with smaller semi-major axes ($a_o$) are more
likely to drive the planet to within $a = 0.01$ au. The bottom panel shows that outer companions with large
eccentricities are more likely to perturb inner planets into very-close-in orbits, but that it is still possible
for outer perturbers with low eccentricities.}
\label{fig:aout_eout}
\end{center}
\end{figure}

Figure \ref{fig:aout_eout} shows how the final semimajor axis of the planet, $a$, depends on the 
orbital properties of the outer body.  The top panel shows that it is more likely that 
the planet will end up close to the parent star, if the outer body 
is in a relatively close orbit ($a_o \sim < 100$ au). \emph{Kepler's} has a 4" pixel size and so 
Figure \ref{fig:mass_a} does suggest that a sufficiently faint, non-variable companion - that could have pertubed a planet
into a Kepler-78b-like orbit - could indeed have avoided detection.  The bottom panel of Figure \ref{fig:aout_eout} shows how 
the final semimajor axis of the planet depends on the on the eccentricity ($e_o$) of the outer body.  
It indicates that close-in orbits are a little more likely when the companion has a high eccentricity ($e > 0.4$) but are 
still possible for those with smaller eccentricities.


\subsection{The pile-up inside $0.1$ au}
Figures \ref{fig:a_final_log_tmax8d8} and \ref{fig:a_final_log_tmax2d9} show a pile-up of planets inside $a = 0.1$ au,
peaking at $\sim 0.02$ au. In our simulations, between $350$ and $850$ (between $3.5$\% and $8.5$\% of the full sample of 10000)
had final semimajor axes inside $a = 0.1$ au (and had
not reached their Roche limit).  If we assume that $40$\% of the \emph{Kepler} sample could have a binary companion (either primordial
or through an exchange interaction) - and that most Sun-like stars have terrestrial-mass,
planetary companions \citep{cassan12,greaves11} - then our results suggest that as much as $3$ \% of the \emph{Kepler} sample might 
have planets that have been perturbed into close-in orbits, with a distribution that peaks at about $0.02$ au. This is intriguingly similar
to the suggestion in \citet{sanchis14} that about 1 in 200 \emph{Kepler} stars hosts a planet with a period of $1$ day or less. 

\subsection{Obliquity}
\begin{figure}
\begin{center}
\includegraphics[scale = 0.435]{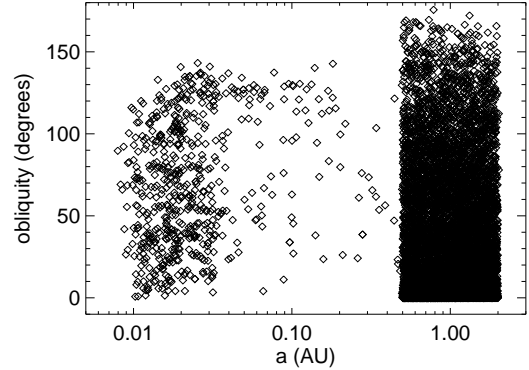}
\caption{Figure showing the obliquity of the inner system.  The inner system starts with the orbital angular momentum
aligned with the spins of the central star and planet.  The perturbation from the outer companion can, however,
cause the inclination of the inner orbit to oscillate and those system that are tidally circularised can end up
with a range of obliquities.}
\label{fig:aplan_inc_inner}
\end{center}
\end{figure}

An interesting aspect of the Kozai-Lidov process is that it can perturb a planet into an orbit that is inclined with respect to its initial
plane and, hence, inclined with respect to the spin of the host star \citep{wu07,fabrycky07}. We now have a number of
close-in, `hot' Jupiters that are on orbits inclined with respect to the spin of the host star \citep{hebrard08, triaud10} and these are
thought to be a consequence of Kozai-Lidov cycles. Figure \ref{fig:aplan_inc_inner} shows the final angle (obliquity) between the angular momentum 
vector of the inner planet's orbit and the spin of the central star, and shows that a wide range of obliquities
are possible.  All the systems
initially have obliquities of zero (the angular momentum of the inner orbit is aligned with the spins of the
parent star and planet) and Figure
\ref{fig:aplan_inc_inner} shows that those that are perturbed into an inner orbit can then be tidally circularised with a large range
of obliquities, consistent with other similar studies \citep{fabrycky07,naoz12}. Using the Rossiter-McLaughlin method to determine such a 
mis-alignment (e.g. \citealt{queloz00}) is probably not possible for such a low-mass planet, but it may be possible to do so using spot-crossing \citep{desert11} or
astro-seismology \citep{chaplin13}.

\section{Discussion and Conclusions}
We have considered, here, if systems like Kepler-78b (an Earth-like exoplanet with a very close-in orbit) could
be due to a perturbation from an outer companion on an, initially, inclined orbit.  To do this, we 
consider a system in which the star and planet have the same masses and radii as in the Kepler-78 system
\citep{sanchis13,pepe13,howard13}, but in which the planet initially has an almost circular orbit with a semi-major axis between $0.5$
and $2$ au.  The system is also assumed to have an outer companion with a semimajor axis between $40$ and $20000$ 
au, with an orbital eccentricity that can be as high as $e = 1$ (but constrained by stability criteria) and that may be 
inclined with respect to the plane of the inner orbit.

We ran a suite of Monte Carlo simulations in which we randomly select the inner and outer systems semi-major axes, 
the eccentricity of the outer system, the mass of the outer companion and the mutual inclination of the two orbits.
We ran two sets of simulations, one where each system was evolved for $t = 800$ Myr, similar to the expected age of the Kepler-78 system,
and the other where the age of the system was randomly selected to be between $500$ Myr and $2$ Gyr.  Our basic results are that :
\begin{itemize}
\item{it is possible for a planet to be perturbed into an orbit similar to that of Kepler-78b around a star with an age ($600 - 900$ Myr) 
similar to that of Kepler-78.  Out of a sample of 10000, between $4$ and $10$ survive inside $a = 0.01$ au.} 
\item{if we consider a broader age range, the likely binarity of the {\em Kepler} sample, and the size of the {\em Kepler} sample,
our results suggest that as many as $20$ of the {\em Kepler} targets with ages less than $2$ Gyr could host a Kepler-78b-like planet. 
Additionally, we find that a system with an age similar to that of Kepler-78 could indeed have been found to host a Kepler-78b-like planet.}
\item{a planet such as Kepler-78b will, quite quickly, reach its Roche limit and be tidally destroyed. Our results suggest
that such a planet would only survive inside $a = 0.01$ au for a few hundred Myrs, at most. In the simulations here, typically
in excess of $100$, but no more than $300$, (out of $10000$) reached their Roche limit and were assumed to be destroyed. This appears consistent with
other work that has also suggested that this process could lead to the tidal destruction of perturbed planets \citep{naoz12}.}
\item{given \emph{Kepler's} 4" pixel size and magnitude limit, it is possible that a faint, non-variable companion that could drive 
Kozai-Lidov cycles may have gone undetected.  That the companion appears to need to be inside $100$ au, means that it may
be possible to detect the resulting radial velocity drift.}
\item{If a planet such as Kepler-78b were perturbed into its current orbit through Kozai-Lidov cycles, we might
expect the star's rotation axis to be misaligned with respect to the planet's orbit. Measuring the star's 
obliquity is quite difficult, but could be possible using spot-crossing \citep{desert11} or using astro-seismology \citep{chaplin13}.}
\item{even though it appears possible that a system such as Kepler-78b could form in this way, it appears to be
more likely for system older than $1$ Gyr, than for systems younger than $1$ Gyr.}
\end{itemize} 

Our basic results, therefore, suggest that such a process could operate, but there are some caveats. 
Although it is possible for a system with an age similar to that of Kepler-78 to host a planet like Kepler-78b, the numbers are small 
(we may expect the {\em Kepler} sample to host only a few such planets). Additionally it seems that it would have been more likely to have found such
a planet in a slightly older system.  These results, therefore, suggest that Kozai-Lidov cycles could have played 
a role in the evolution of Kepler-78b, but don't rule out that there could be an alternative explanation, such as planet-planet scattering \citep{rasio96}.This, however, may suffer from the similar issues, since the dominant constraint - given the relatively low age of Kepler-78 - is the tidal evolution timescale. 
This constraint would probably also apply to the tidal downsizing hypothesis \citep{nayakshin10},
in which a massive gas-giant planet formed in the outer parts of the system, migrates rapidly inwards and loses masses via tidal stripping.
That, of course, leaves the possibility that disc migration \citep{ward97} moved this planet into a very close orbit which has since evolved,
through tidal interactions with the host star, into the orbit it inhabits today.  Again, this would also involve tidal evolution once
the disc has dispersed and so may also have a similar timescale issue, unless disc migration can place the planet sufficiently close to the
parent star so that it can then tidally evolve to where it is today.   

Recent work by \citet{sanchis14} suggest that about 1 out of every 200 stars hosts an ultra short period (USP) planet (period of 1 day or less).  Although we've focussed on Kepler-78b here and found that few of our simulated systems have final periods as short as Kepler-78b (8.5 hours), many more have periods
of $1$ day or less. The exact number depends on the chosen parameters, but it varies from $\sim 100$ to just over $300$ (from a total sample
of 10000).  Given that not all stars are binaries, this
is intriguingly similar to the result in \citet{sanchis14}. Similarly, our results suggest that such a process should lead to a pile-up 
of planets with a peak at about $0.02$ au, again consistent with \citet{sanchis14} who find that the occurence rate rises with 
period from 0.2 to 1 day.   It may, however, be difficult to distinguish a pile-up due to Kozai-Lidov cycles from what is 
expected from scattering in multi-planet systems \cite{schlaufman10}.  \citet{sanchis14} do, however, suggest that
almost all USPs have companion planets with period $P < 50$ days, which may provide a constraint on the formation process for these USPs.

We should also acknowledge the possibility that our assumptions do not properly represent the possible initial conditions such a system
could have. The initial distribution of the planet in semimajor axis space may be different to what we've assumed and the
orbital properties of the outer perturber may also be different.  Similarly, the tidal properties of the parent star and
planet may differ from what we've assumed. However, we should at least acknowledge that even though our results suggest that
Kozai-Lidov cycles will rarely produce a planet with properties similar to that of Kepler-78b, Kepler-78b is itself rare.  In that
sense our results could be seen as somewhat consistent with our knowledge of such planets, but that - alone - doesn't allow
us to determine if it is likely that such a process did indeed play a role in the evolution of Kepler-78b.

\noindent 

\section*{Acknowledgements}
The author acknowledges very useful discussions with Adrian Barker, Gordon Ogilvie, Douglas Heggie, Andrew
Collier Cameron, Duncan Forgan and Eric Lopez. KR gratefully acknowledges support from STFC grant ST/J001422/1. The author would also 
like to thank the anonymous referee whose comments substantially improved this paper. The research leading to these results
also received funding from the European Union Seventh Framework Programme (FP7/2007-2013) under grant agreement number
313014 (ETAEARTH).

\label{lastpage}

\end{document}